\documentclass[11pt]{article}

\usepackage{amsmath,amssymb}

\newcommand{\proof}{ $\triangleright$\quad}
\newcommand{\qed}{\hfill $\triangleleft$}

\begin{document}

\title{Bosonization method for second  super quantization}
\author{Alexander  Dynin\\
\textit {\small Department of Mathematics, Ohio State University}\\
\textit {\small Columbus, OH 43210, USA}, \texttt{\small dynin@math.ohio-state.edu}}
 
 \maketitle

\begin{abstract}
 A bosonic-fermionic correspondence allows an analytic definition of functional  super derivative, in particular,  and a bosonic functional calculus, in general, on
  Bargmann- Gelfand  triples for the second super quantization. A Feynman integral for the super transformation matrix elements in terms of bosonic anti-normal  Berezin symbols is  rigorously constructed. 
\end{abstract}

\medskip



\begin{flushright}
In memoriam of F. A. Berezin (1931-1980).
\end{flushright}
s

\section{Introduction}
\subsection{Preview}
 \begin{itemize}
 
    \item    We begin with a summary of research directions opened   by Berezin's 
monograph  "Method of second quantization"  \textsc{berezin}\cite{Berezin-87}:
one is a super second quantization, and another is an extension of   quantum mechanical Schr\"{o}dinger  picture to quantum field theory.

    \item  
 Extending \textsc{kree}\cite{Kree-77}'s second quantization and Hida's white noise calculus (see, e.g., \textsc{obata}\cite{Obata-94}) we develop  the second quantization  in  super Bargmann Fock Gelfand triples  to account for   the quantum states knocked out by a violent Schr\"{o}dinger operator (see below the  quote from Dirac.) 

The "violence" means  that the domain  of a
 Schr\"{o}dinger operator is not dense in  the Fock space. Actually, it  is   continuous operator from a nuclear Frechet space of test functionals of classical fields to the  anti-dual space of functional  of distributions. (Apparitions of Gelfand triples  are seen in \textsc{berezin}\cite[Subsections II.2.5 and II.3.4)]{Berezin-65}).

\item The  related $c\hbar$ diffculty in \emph{relativistic} quantum field theory   was discovered by \textsc{landau-peierls}\cite{Landau-31} in 1931:  boundedness of velocities  by the light velocity $c$ implies boundedness of momenta, so that, by the uncertainty principle, exact  values of a  quantized field do not exist. However  the difficulty is resolved  in Gelfand triples via canonical commutation relations between creation operators of  test functionals and annihilation operators of distribution functionals. 

  \item  
  Following   \textsc{dynin}\cite{Dynin-09},  Schr\"{o}dinger super equations  in a Gelfand triple are solved via  mathematically rigorous \emph{anti-normal} Feynman super integral.
 \end{itemize}

\subsection{Berezin's legacy}
  In 1956 F. Berezin was initiated into Quantum Field theory by  I. M. Gelfand. He was  greatly influenced by  K. Friedrichs' dictum:
\begin{quote}
According to Niels Bohr, any attempt at a sharp definition of physical concepts
must even violate their real physical meaning. Therefore, the mathematician's 
desire for a deductive presentation of physical theory cannot be established in principle. On the other hand, it seems justified to strive for a precise definition of the intrinsic mathematical meaning of mathematical notions employed in [...] quantum theory. \ (\textsc{friedrichs}\cite{Friedrichs-53})
\end{quote} 
   Berezin's goal was
\begin{quote}
to construct a noncontradictory quantum field theory. Without exaggeration, it can be said that  almost all of his work (on the $N$ particle problem, quantization, superanalysis) he regarded as stepping stones to this difficult problem. \ (\textsc{minlos}\cite{Minlos-07})
\end{quote}
In this respect Berezin's monograph  \textsc{berezin}\cite{Berezin-65} became the next landmark after  \textsc{friedrichs}\cite{Friedrichs-53}.
It  develops  the Fock method of generating functionals of bosonic states into a  correspondence between quantum operators and their normal functional symbols in the framework of (anti-)holomorphic Fock spaces of  bosonic states and, for the first time,  of  fermionic states.  The normal representation of bounded operators in  Fock spaces came as a surprise and, certainly, asked for further generalizations.  That was done in  bosonic Gelfand triples of quantum field theory \textsc{kree}\cite{Kree-77} and of white noise calculus (see \textsc{obata}\cite{Obata-94}). For fermions this is done in this paper.

 The following up Berezin's  papers   on quantization with finite number of freedom degrees have been stepping stones toward quantum field theory.   In particular, analytical  possibilities of anti-normal  Sudarshan, or diagonal) 
 (aka  Berezin,  Hisumi, Sudarshan, diagonal)  symbols have been explored in \textsc{berezin}\cite{Berezin-71}.

Parallelism between bosonic and fermionic Fock spaces was was already discussed in  \textsc{fock}\cite{Fock-32}.  In the bosonic case the functional method of the second quantization was proposed in the sequel \textsc{fock}\cite{Fock-34}.   The "striking similarity" of   \emph{analysis}  of bosonic and fermionic  generating functionals (\textsc{berezin}\cite[Introduction]{Berezin-65}) inspired Berezin's fermionic extension of the functional method. (By penetrating remark in \textsc{neretin}\cite[Section I.5]{Neretin}, \textit{for this it was necessary to understand well both bosonic and fermionic cases. Otherwise they are not so similar...}.)
Eventually, this led to   super  analysis (cp.  \textsc{berezin}\cite{Berezin-87}) and to new beginnings in theory of infinite linear groups (cp.  \textsc{neretin}\cite{Neretin}). 

\smallskip 
By \textsc{berezin}\cite[Introduction]{Berezin-65}, 
\begin{quotation}
[...] functionals may be imagined, roughly speaking, as functions of infinitely many variables. In usual quantum mechanics, the number of variables of functions representing  the state space is the number of freedom degrees.
 
 Thus there arises an interpretation of the second quantization problems  
as quantum mechanics problems with infinitely many degrees of freedom and a natural desire  to approximate  these problems via problems with finite, but large, number of degrees of freedom.
\end{quotation} 
In the last section of this paper we use such interpretation
to derive a rigorous anti-normal Feynman integral for the  matrix elements
Schr\"{o}dinger operators in super Gelfand triples.

\subsection{Quantum mechanics vs quantum field theory}
Quantum Mechanics was invented by  W. Heisenberg in 1925. The  famous monographs von \textsc{neumann}\cite{Neumann-32} and \textsc{weyl}\cite{Weyl-31} summarized corresponding  new mathematics. The main goal was to comprehend canonical commutation relations  and ensuing non-commutativity of quantum variables of Heisenberg's  and E. Shr\"{o}dinger's Quantum Mechanics.

von Neumann  defined and named Hilbert spaces to honor Hilbert theory of quadratic forms. He replaced the latter by  (unbounded) self-adjoint operators 
corresponding to quantum observables.  Weyl    quantization converts  classical  observables into operators. A generalized  quantization rule was proposed by \textsc{wigner}\cite{Wigner-32} and the corresponding formal calculus was developed by \textsc{agarwal-wolf}\cite{Agarwal-70}.

The  fundamental quantum uncertainty principle has the mathematical
underpinning  of canonical commutation relations.
Weyl conjectured  and von Neumann  proved the unitary equivalence  of  irreducible unitary  representations of bosonic canonical commutation relations with a finite number of degrees of fredom.
The corresponding theorem for fermionic  canonical commutation relations was 
established in 1927 by P. Jordan and E. Wigner.

\smallskip

In contrast to quantum Mechanics, mathematics of Quantum Field Theory has been developing much slower, mainly because of   the infinite number  of degrees of freedom.  There was  vivid correspondence between W. Heisenberg, P. Jordan, and W. Pauli about possibilities of  Volterra  functional calculus. 

\smallskip
There is no uniqueness theorem for unitary 
representations of the canonical commutation relations (cp. the monograph
\textsc{friedrichs}\cite{Friedrichs-53}) (it is presumed that this non-umniquness was discovered by von Neumann in late 1930's). However, under an additional requirement of existence of the fiducial quantum vacuum state, they  are unitarily equivalent. 

\textsc{friedrichs}\cite{ Friedrichs-53} was  an attempt  to do for  Quantum Field Theory what von   Neumann had done for Quantum Mechanics. Unfortunately, it lacked von Neumann elegance. 
F. Berezin's use of holomorphic Fock-Bargmann  representations in his monograph \textsc{berezin}\cite{Berezin-65} clarifies  the subject  considerably.

\medskip
In 1927  \textsc{dirac}\cite{Dirac-27} introduced the method of second quantization in Quantum Electrodynamics as a system of harmonic oscillators.

In  1931  \textsc{landau-peierls}\cite{Landau-30} proposed  an alternative   configuration space quantization method  diagonalizing the field multiplication operators. ( Monograph \textsc{berezin}\cite{Berezin-65} starts with  such configuration space,  no attribution already needed.)
 
The foundational  paper \textsc{fock}\cite{Fock-32} on \emph{Fock representation} of the bosonic \emph{and} fermionic canonical commutation relations  with  the infinite number  of degrees of freedom begins as follows (in translation)
\begin{quote}
The fact that the second quantization method is equivalent to the method of usual wave functions on a  configuration space is known in principle. In this paper this is traced in detail.
\end{quote}
In the presence of a unique vacuum vector the representations of canonical commutation relations are  unique up to unitary equivalence. 

The sequel \textsc{fock}\cite{Fock-34} introduced the method of generating functionals for \emph{bosons}.    Both Fock papers are formal calculations.

\medskip 
In the beginning, W. Heisenberg, P. Jordan, and W. Pauli had idea to   use  the canonical commutation relations to extend  Heisenberg and  Shr\"{o}dinger  pictures of quantum mechanics to quantum field theory. There was a vivid discussion of "Volterra mathematics"  in their correspondence.

However,  according to  P. Dirac ( ``Lectures on quantum field theory" Yeshiva University, N.Y. 1966, 
Section  ``Relationship of the Heisenberg and  Schr\"{o}dinger Pictures"),
\begin{quotation}
The interactions that are physically important in quantum field theory are so violent 
that they will knock  any Schr\"{o}dinger state vector out of Hilbert space  
in the shortest possible time interval.

[...] It is better to  abandon all attempts at using the Schr\"{o}dinger picture 
with these Hamiltonians.

[...] I don't want to assert that the Schr\"{o}dinger picture will not come back. 
In fact, there are so many beautiful things about it that I have the feeling in 
the back of my mind that it ought to come back. I am really loath to have to give it up.
\end{quotation}
 In this paper Schr\"{o}dinger picture is resurrected in the framework of Gelfand triples, cp. \textsc{gelfand-vilenkin}\cite{Gelfand}.

\section{Bosonic Gelfand triples} 
\subsection{Holomorphic states}
In this section, $\mathcal{H}$ denotes an (infinite dimensional)  \emph{bosonic} complex separable Hilbert ${*}$-space  with conjugation (cp. \textsc{berezin}\cite{Berezin-65}). 

Sandwich   $\mathcal{H}$   into   a Gelfand nuclear  ${*}$-triple (cp., \textsc{gelfand-vilenkin}\cite{Gelfand})
\begin{equation}
\label{ }
\mathcal{H}^\infty\subset\mathcal{H}\subset\mathcal{H}^{-\infty},
\end{equation}
where  $\mathcal{H}^\infty$ is a nuclear countably Hilbert  ${*}$-space,  
$\mathcal{H}^{-\infty}$ is its topological ${*}$-dual with respect  to the Hermitian product on $\mathcal{H}$. The  imeddings are continuous  with dense ranges and real, i. e., commute with the conjugation.
 
 By   Minlos' theorem, space $\mathcal{H}^{-\infty}$ carries the probability Gauss Radon measure   $dz^*dz\:e^{-z^*z}$.
This symbolic expression is meaningful  as a  cylindrical measure on
 $\mathcal{H}^{-\infty}$ which extends to the Gauss-Radon measure. We use the same notation for both measures because it allows integration by parts and Fubini theorem   which hold for integrals of cylindrical functions followed by limit transition to the wider class of integrable functions.

 Fernique's theorem (see \textsc{bogachev}\cite[Chapter 2]{Bogachev})  implies that there exists a positive 
constant $c$ such that if a functional $\Psi(z^*)$ on $\mathcal{H}^{-\infty}$ 
is continuous   and $\Psi \prec e^{-cz^*z}$ then  $\Psi(z^*)$ is integrable on
 $\mathcal{H}$.

\medskip
The \emph{Bargmann space} (see, e. g., \textsc{berezin} \cite[Chapter I]{Berezin-65}) is the (complete) complex Hilbert space of \emph{G\^{a}teaux entire} functionals $\Psi=\Psi(z^*)$ on 
 $\mathcal{H}^{-\infty}$ with conjugation
\begin{equation}
\label{eq:conj}
\Psi^*=\Psi^*(z) \equiv\ \overline{\Psi(z^*)}
\end{equation}
and  integrable  Hermitian  sesqui-lnear inner  product 
\begin{equation}
\Psi^*\Phi\ \equiv \int\!dz^*dz\: e^{-z^*z}\Psi^*(z)\Phi(z^*).
\end{equation}
The  integral is is  denoted also as the \emph{Gaussian contraction}
 $\Psi^*(z)\Phi(z^*)$.

The \emph{exponential functionals} 
\begin{equation}
e^{z}(z^*) \equiv e^{z^*z},\quad z\in \mathcal{H}^\infty,
\end{equation}
belong to $\mathcal{B}^0$ since $e^{z^*}e^{z}=e^{z^*z}<\infty$. Indeed  
\begin{equation}
e^{z^*}e^{\xi}\ =\ \int\!dz^*dz\: e^{-z^*z}e^{z^*z+z^*\xi}\ =\ 
e^{z^*\xi}\int\!dz^*dz\: e^{-(z^*-z^*)(z-\xi)}\ =\ e^{z^*\xi}
\end{equation}
They form  a  \emph{continual orthogonal basis} of exponential functionals in
$\mathcal{B}^0$ (see, e. g., \textsc{berezin} \cite[Chapter I]{Berezin-65}): 
If $\Psi=\Psi(z^*)\in\mathcal{B}^0$ then the  \emph{Borel transform}
\begin{equation}
\Psi(z^*)\ =\  e^{-z^*z}\int\!d\zeta d\zeta^* e^{-\zeta^{*}\zeta}\widetilde{\Psi}(\zeta) e^{z^*\zeta},
\quad \ \widetilde{\Psi}(\zeta)\ \equiv\ \Psi^*e^{\zeta}.
\end{equation}
is a unitary operator in $\mathcal{B}^0$.  

The orthogonal basis  is overcomplete  since 
\begin{equation}
e^{z}\ =\  \int\!d\zeta d\zeta^* e^{-\zeta^*\zeta}  e^{z^*\zeta}.
\end{equation}

\medskip
 \emph{Bargmann-Hida space} $\mathcal{B}^\infty$ is the vector space of  of G\^{a}teaux entire  \emph{test   functionals}  $\Psi(z^*)$  on  $\mathcal{H}^{-\infty}$  of the (topological)  second order and minimal type, i. e., for any $s\geq 0$ and  $\epsilon>0$ there exists a constant  $C >0$ such that\begin{equation}
|\Psi(z^*)| \leq \, Ce^{\epsilon \|z^*\|_{-s}^2},\quad z^*\in  \mathcal{H}^{-s}.
\end{equation}
 $\mathcal{B}^\infty$ is a nuclear countably Hilbert space
  dense in $\mathcal{B}^0$    (see, e. g., \textsc{obata} \cite[Section 3.6]{Obata-94}).
   
   Actually, the topology of $\mathcal{B}^\infty$ is defined by the norms
\begin{equation}
\| \Psi \|_{s,\epsilon}\ \equiv\ \sup_{z^*} |\Psi(z^*)| e^{-\epsilon \|z^*\|_{-s}^2}.
\end{equation}
Again, by \textsc{obata} \cite[Section 3.6]{Obata-94}), Borel transform is a topological automorphism
of $\mathcal{B}^\infty$.
 \smallskip
\emph{Bargmann-Hida space} $\mathcal{B}^{-\infty}$ of \emph{generalized functionals} 
$\Psi^*(z)$ on  $\mathcal{H}^{\infty}$ is the strong anti-dual space  of 
 $\mathcal{B}^\infty$.

  The   Borel transform
 $\widetilde{\Psi}^*(z)$ of  $\mathcal{B}^{-\infty}$ is defined as the anti-dual  of the  Borel transform of  $\mathcal{B}^\infty$ of $\mathcal{B}^{-\infty}$ (and, therefore, a topological automorphism).
 
  By (e.  g., \textsc{obata},  \cite[Section 3.6]{Obata-94} , the generalized functionals  are characterized as entire functionals of the (bornological) second order  on $\mathcal{H}^{\infty}$, i. e.,  there exist positive constants $C, K$  and $s\geq 0$ such that 
\begin{equation}
|\Psi(z)| \leq \, Ce^{K\|z\|_{s}^2},\quad z\in \mathcal{H}^{s}.
\end{equation}
We get the Bargmann-Hida Gelfand triple of holomorphic states
\begin{equation}
\mathcal{B}^\infty\ \subset\ \mathcal{B}^0\ \subset \mathcal{B}^{-\infty}.
\end{equation}
The vector spaces  $\mathcal{B}^{\infty}$  and  $\mathcal{B}^{-\infty}$  are locally convex commutative topological algebras with the point-wise multiplication. Moreover we have Taylor expansions
\begin{eqnarray}
\label{eq:entire}
& &
\Psi(z^*+w^*)=\sum_{n=0}^\infty \frac{\partial_{z^*}^n\Psi(z^*)}{n!}w^{*n}\quad \mbox{for}\
\Psi\in\mathcal{B}^{\infty},\\ 
& &
\Psi(z+w)=\sum_{n=0}^\infty \frac{\partial_{z}^n\Psi(z)}{n!}w^{n}\quad \mbox{for}\
\Psi\in\mathcal{B}^{-\infty}.
\end{eqnarray}

\medskip
By conjugating   $z$ to  $z^*$, we convert  $\mathcal{H}^\infty\ \subset\ \mathcal{H}^0\ \subset\ \mathcal{H}^{-\infty}$ into the conjugate Gelfand triple
${H}^{*\infty}\ \subset\ {H}^{*0}\ \subset\ 
{H}^{*-\infty}$. Their direct product
\begin{equation}\label{eq:product}
\mathcal{H}^{\infty}\times {H}^{*\infty}\ \subset\ \mathcal{H}^{0}  \times {H}^{*0}\ \subset\ 
{H}^{-\infty}\times \mathcal{H}^{*-\infty}
\end{equation}
carries the complex conjugation $ (z,w^*)^*\equiv (w,z^*)$.

 The Bargmann-Hida   Gelfand triple associated with (\ref{eq:product})  is  \begin{equation}
(\mathcal{B}\otimes\mathcal{B}^*)^\infty\ \subset\ (\mathcal{B}\otimes\mathcal{B}^*)^0\ \subset (\mathcal{B}\otimes\mathcal{B}^*)^{-\infty}. 
 \end{equation} 
 Entire functionals  $\Theta(z,w^*)\in(\mathcal{B}\otimes\mathcal{B}^*)^{-\infty}$ are uniquely 
 defined  by    their restrictions $\Theta(z,z^*)$ to the real diagonal.  If   
 $\overline{\Theta(z,z^*)} = \Theta(z,z^*)$, then $\Theta(z,z^*)$ is a \emph{classical observable} on the phase space $\mathcal{H}^\infty$. 
 
By  Bargmann-Segal transform (see, e. g., \textsc{obata}\cite[$S$-transform]{Obata-94}),  Cook-Fock Gelfand $*$-triple $\mathbf{F}$ is unitarily equivalent to Bargmann-Hida Gelfand $*$-triple $\mathbf{B}$. 

\subsection{Second quantization of classical bosonic observables}

For $z\in\mathcal{H}^\infty,\  z^*\in\overline{\mathbb{C}}\mathcal{H}_0^{-\infty}$ 
define four continuos operators of multiplication 
and directional complex differentiation (\emph{operators of creation and annihilation}):
\begin{eqnarray}
\label{eq:creator}
\hat{z}: \mathcal{B}^{\infty}\rightarrow \mathcal{B}^{\infty},\quad 
& &
\hat{z}\Psi(\zeta^*)\ \equiv\ z \Psi(\zeta^*)\ =\  (\zeta^*z)\Psi(\zeta^*);\\
\hat{z}^\dagger: \mathcal{B}^{-\infty}\rightarrow \mathcal{B}^{-\infty},
 & &
\hat{z}^\dagger\Psi(\zeta)\ \equiv\ \partial_z\Psi(\zeta);\\
\widehat{z^*}: \mathcal{B}^{-\infty}\rightarrow \mathcal{B}^{-\infty},
 & &
\widehat{z^*}\Psi(\zeta)\ \equiv\ z^*\Psi(\zeta)\ =\ (z^*\zeta)\Psi(\zeta);\\
\label{eq:annih}
\widehat{z^*}^\dagger: \mathcal{B}^{\infty}\rightarrow \mathcal{B}^{\infty},
 & &
\widehat{z^*}^\dagger\Psi(\zeta^*)\ \equiv\ \partial_{z^*} \Psi(\zeta^*).
\end{eqnarray}
\proof
The continuity of multiplications is straightforward  and of directional differentiations is
by Cauchy integral formula for the derivative of a holomorphic function. \qed 

These operators generate strongly continuous abelian operator groups in 
$ \mathcal{B}^{\infty}$ and $ \mathcal{B}^{-\infty}$:
\begin{eqnarray}
e^{\hat{z}}: \mathcal{B}^{\infty}\rightarrow \mathcal{B}^{\infty},\quad & &
e^{\hat{z}}\Psi(\zeta^*)\ =\ e^{\zeta^*z}\Psi(\zeta^*);\\
e^{\hat{z}^\dagger}: \mathcal{B}^{-\infty}\rightarrow \mathcal{B}^{-\infty},\quad & &
e^{\hat{z}^\dagger}\Psi(\zeta)\ =\Psi(\zeta+z);\\
e^{\widehat{z^*}}: \mathcal{B}^{-\infty}\rightarrow \mathcal{B}^{-\infty},\quad & &
e^{\widehat{z^*}}\Psi(\zeta)\ =\ e^{z^*\zeta}\Psi(\zeta);\\
e^{\widehat{z^*}^\dagger}: \mathcal{B}^{\infty}\rightarrow \mathcal{B}^{\infty},\quad & &
e^{\widehat{z^*}^\dagger}\Psi(\zeta^*)\ =\  \Psi(\zeta^*+z^*).
\end{eqnarray}
The only non-trivial commutator relations for the groups
\begin{equation}\label{eq:expcom}
[e^{\widehat{z^*}^\dagger},  e^{\hat{z}}]\ =\  e^{z^*z},\quad
[e^{\hat{z}^\dagger},  e^{\widehat{z^*}}] \ =\  e^{zz^*}
\end{equation}
imply the only non-trivial commutator relations for their generators
\begin{equation}\label{eq:com}
[\widehat{z^*}^\dagger,\hat{z}]\ =\ z^*z,\quad  [\hat{z}^\dagger,\widehat{z^*}]\ =\ zz^*.
\end{equation}

\medskip
The \emph{normal quantization}  $\Theta(\hat{z},\widehat{z^*}^\dagger)$ of 
$\Theta(z,z^*)\in(\mathcal{B}\otimes\mathcal{B}^*)^{-\infty}$ is defined as  the  continuous linear operator
\begin{equation}
\Theta(\hat{z},\widehat{z^*}^\dagger):\mathcal{B}^{\infty}\rightarrow\mathcal{B}^{-\infty}
\end{equation}
 via the  sesqui-linear quadratic form (in Einstein -DeWitt contraction notation, i. e., in the integral contraction over conjugated \emph{continual} variables the integration symbols  are skipped)
 \begin{equation}
 \label{eq:normal}
\Psi^*(z)\Theta(\hat{z},\widehat{z^*}^\dagger)\Psi(z^*)\ \equiv\  
\widetilde{\Theta}(\zeta^*,\zeta)e^{\hat{z}}
e^{\widehat{z^*}^\dagger}\widetilde{\Psi}^*(\zeta)\widetilde{\Psi}(\zeta^*).
\end{equation}
The sesqui-holomorphic   $\widetilde{\Theta}(\zeta^*,\eta)$ is the \emph{normal symbol} of the operator $\Theta(\hat{z},\widehat{z^*})$ uniquely defined by its restriction
$\widetilde{\Theta}(\zeta^*,\zeta)$ to the real diagonal.

Similarly, the  \emph{kernel} $K(z,y^*)$ of the operator $\Theta(\hat{z},\widehat{z^*})$ is uniquely defined by its diagonal restriction
\begin{eqnarray}
& &
\label{eq:kernel}
K(z,z^*)\ \equiv\ e^{\zeta^*}(z)\Theta(\hat{z},\widehat{z^*}^\dagger)e^{\zeta}(z^*)\\ 
& &
\stackrel{(\ref{eq:normal})}{=}
\widetilde{\Theta}(\zeta^*,\zeta)e^{\hat{z}}
e^{\widehat{z^*}^\dagger}e^{z}(\zeta^*)e^{z^*}(\zeta)\stackrel{(\ref{eq:creator}),(\ref{eq:annih}))}{=}\Theta(z,z^*)e^{zz^*}
\in(\mathcal{B}\otimes\mathcal{B}^*)^{-\infty}
\end{eqnarray}
 Thus the correspondence between \emph{quantum observables} 
 $\Theta(\hat{z},\widehat{z^*})$ and classical observables  $\Theta(z,z^*)$   is one-to-one:
\begin{equation}\label{eq:antinormal}
K(z,z^*)\ =\ \Theta(z,z^*)e^{zz^*}.
\end{equation}
SInce  $e^{zz^*}$ is the integral kernel of the orthogonal projection of
$(\mathcal{B}\otimes\mathcal{B}^*)^{-\infty}$ onto 
$\mathcal{B}^{*\infty}$,
the classical variable $\Theta(z^*,z)$ is the \emph{Berezin (aka antinormal, diagonal, or Sudarshan) symbol} of the operator $\Theta(\hat{z},\widehat{z^*}^\dagger)$, i.e., the compression  of the multiplication  with $\Theta(z^*,z)$ to  $\mathcal{B}^{*\infty}$:
\begin{equation}
\label{eq:diagonal}
\Theta(\hat{z},\widehat{z^*}^\dagger)\Psi(z^*)\ =\ e^{z^*z}\Theta(z^*,z)\Psi^*(z). 
\end{equation}
The symbol is called antinormal because  
\begin{equation}\label{l}
\Psi^*(z)e^{zz^*}\Theta(z^*,z)\Psi(z^*)\ \stackrel{(\ref{eq:expcom})}{=}\ 
\widetilde{\Psi}^*(\zeta^*)\widetilde{\Theta}(\zeta^*,\zeta)
e^{\widehat{z^*}^\dagger}e^{\hat{z}}\widetilde{\Psi}(\zeta).
\end{equation}
Compare with   the (opposite) normal operator ordering in (\ref{eq:normal}).

\smallskip
For $\Theta\in(\mathcal{B}\otimes\mathcal{B}^*)^{\infty}$ we have, 
by Taylor expansion and integration by parts,
\begin{eqnarray*}
& &
\Theta(z,z^*) = e^{-z^*z}\tilde{\Theta}(\zeta^*,\zeta)e^{\zeta^*z}e^{z^*\zeta} 
\\ & &
=\ \int\! d\zeta^*d\zeta\: \tilde{\Theta}(\zeta^*,\zeta)e^{-(z^*-\zeta^*)(z-\zeta)}
= \int\! d\zeta^*d\zeta\: e^{-\zeta^*\zeta} \tilde{\Theta}(z^*-\zeta^*,z-\zeta)
\\ & &
=\ \sum_{k,m}\frac{(-1)^{k+m}}{k\! !\: m\! !}\int\! d\zeta^*d\zeta\: e^{-\zeta^*\zeta} 
\partial_{\zeta^*}^k \partial_\zeta^m{\tilde{\Theta}}(z^*,z)(\zeta^{*k}\zeta^m)
\\ & &
=\ \sum_{k}\frac{1}{k\! !}\partial_{\zeta^*}^k \partial_\zeta^m\tilde{\Theta}(z^*,z)
\int\! d\zeta^*d\zeta\: e^{-\zeta^*\zeta}(\zeta^{*k}\zeta^m) \ =\ 
e^{(1/2)\partial_{\zeta^*} \partial_\zeta}\tilde{\Theta}(z^*,z),
\end{eqnarray*}
since $\widehat{\zeta^*}^\dagger= \partial_\zeta$. (Note the contraction 
$\partial_{\zeta^*} \partial_\zeta$ is an infinite dimensional  complex Lapacian.)

Since  $(\mathcal{B}\otimes\mathcal{B}^*)^{\infty}$ is dense in $(\mathcal{B}\otimes\mathcal{B}^*)^{-\infty}$ we get  the
relationship  between the normal and antinormal symbols for all 
$\Theta\in(\mathcal{B}\otimes\mathcal{B}^*)^{-\infty}$ as
\begin{equation}
\label{eq:heat}
\Theta(z^*,z)\ =\ e^{(1/2)\partial_{\zeta^*} \partial_\zeta}\tilde{\Theta}(z^*,z)\quad \mbox{and}\quad
\tilde{\Theta}(z^*,z)\ = e^{-(1/2)\partial_{\zeta^*} \partial_\zeta}\Theta(z,z^*).
\end{equation}
E.g., the constant functional $1$ is both  the normal and anti-normal symbols of the identity operator; the normal symbol of the number operator is the functional
$z^*z$, and its anti-normal symbol is $z^*z-1/2$.
\section{Gelfand super triples}
\subsection{Bosonization of Gelfand super triples}
Consider an infinite dimensional (separable)  complex  super  Hilbert $*$-space 
$\mathcal{H}$, i. e., a  $\mathbb{Z}^2$-graded space 
\begin{equation}
\label{ }
\mathcal{H}\ =\ \mathcal{H}_0\  \oplus \mathcal{H}_1, \quad \mbox{dim}(\mathcal{H}_1)\ =\ \infty
\end{equation}
of elements $z=z_0\oplus z_1$ with bosonic ezen parts $z_0$    and   the  fermionic odd parts $z_1$.  we denote the parity of a homogeneous  element $z$ by $p(z)$. 
Furthermore, $\mathcal{H}$ is endowed with   an anti-linear even  involution 
 $*:\ z\mapsto z^*$, i.e.,
  \begin{equation}
\label{ }
 (cz)^*\ =\ \bar{c}z^*, c\in\mathbb{C},\quad z^{**}\ =\ z.
 \end{equation}
  The super Hermitian product $z^*w$ on $\mathcal{H}$ is a super sesqui-linear form  on $\mathcal{H}$ such that for homogeneous $z,w\in\mathcal{H}$ \begin{equation}
\label{ }
\overline{z^*w}\ =\ (-1)^{p(z)p(w)} w^*z,\quad \ z_0^*z_0\ \geq 0, \quad \ -iz_1^*z_1\geq 0.
\end{equation}
In particular,  the super Hermitian quadratic form  $z_1^*z_1$ is pure imaginary with $\Im (z_1^*z_1)\geq 0$. Moreover, $\mathcal{H}_0$ and $\mathcal{H}_1$ are super orthogonal spaces. 

 Infinite dimensional separable Hilbert $*$-spaces haze real orthonormal bases. Therefore,  they are $*$-unitarily isomorphic, i. e., the equivalence  unitary operator commutes with complex conjugations (cp. \textsc{berezin}\cite[Introduction]{Berezin-65}).
 
\medskip
The infinite dimensional separable super Hilbert $*$-space  $\mathcal{H}_1$   is super untary equivalent to  the Hilbert $*$-space   $\mathcal{L}^2(\mathbf{R})$ of complex-valued functions on the real line with the usual complex conjugation.

Then the anti-symmetric  Hilbert tensor power $\otimes_1^n\mathcal{H}_1$ is 
$*$-unitarily equivalent to  the Hilbert $*$-subspace of anti-symmetric functions
$f(x_1,...,x_n)$ in $\mathcal{L}^2(\mathbf{R}^n)$.

Both symmetric and anti-symmetric functions  on $\mathbf{R}^n$ are uniquely defined by restriction  to the open subset   of $\mathbf{R}^n$
\begin{equation}
\label{ }
\check{\mathbf{R}}^n\ \equiv\  \{\check{x}=(x_1,x_2,...,x_n):\ x_1<x_2<...<x_n\}.
\end{equation}
 The super symmetrization $S_0^n$
of  $f(\check{x})$ on $\check{\mathbf{R}}^n$ produces   a unique super symmetric function $S_0^n(f)(x_1,...,x_n)$ \emph{almost everywhere} on $\mathbf{R}^n$. Furthermore,  $S_0^n$ generates a
$*$-unitary operator  from $\otimes_1^n\mathcal{H}_1$ onto 
$\otimes_0^n\mathcal{H}_1$ (cp., \textsc{meyer}\cite[pp. 59--60]{Meyer}). This implies $*$-\emph{bosonization}  unitary isomorphisms  
\begin{equation}
\label{eq:bosonization}
\varpi^{m,n} :\ \otimes_0^m\mathcal{H}_0\ \oplus\ \otimes_1^n\mathcal{H}_1\
\longrightarrow\  \otimes_0^m\mathcal{H}_0\ \oplus\ \otimes_0^n\mathcal{H}_1.
\end{equation}
The direct sum $\varpi = \oplus_{m,n}\varpi^{m,n}$  is the $*$-unitary bosonization of the super Fock space $\mathcal{F(H)} = \mathcal{H}_0\otimes  \mathcal{H}_1$. It converts $\mathcal{F(H)}$ into the bosonic  Fock space $\mathcal{F(H_1)}$ over the bosonic Hilbert 
$*$-space $H_1=\mathcal{H}_0\oplus\varpi^{0,1}\mathcal{H}_1$.
 The odd mapping  $\varpi$ is  linear and  super unitary:  $\varpi^\dag= \varpi^{-1}$.  

Let  $\mathcal{H}_1$ be sandwiched into the bosonic  Bargmann-Hida Gelfand $*$-triple
\begin{equation}
\label{ }
\mathcal{H}_1^\infty\subset\mathcal{H}_1\subset
\mathcal{H}_1^{-\infty}.
\end{equation}
Then $\mathcal{H}$ is sandwiched into the  Gelfand \emph{super $*$-triple}
\begin{equation}
\label{ }
\mathcal{H}^\infty\subset\mathcal{H}\subset\mathcal{H}^{-\infty},
\end{equation}
where $\mathcal{H}^\infty\equiv \varpi^{-1}(\mathcal{H}_1^\infty)$
is a countably super Hilbert $*$-space, and $\mathcal{H}^{-\infty}$ is its topological $*$-dual.

As a  consequence, the corresponding holomorphic Bargmann-Hida Gelfand bosonic triple
over $\mathcal{H}_1$
\begin{equation}
\mathcal{B}_1^\infty\ \subset\ \mathcal{B}_1^0\ \subset \mathcal{B}_1^{-\infty}.
\end{equation}
 is transformed into the  holomorphic Bargmann-Hida Gelfand \emph{super $*$-triple} over $\mathcal{H}$
\begin{equation}
\mathcal{B}^\infty\ \subset\ \mathcal{B}^0\ \subset \mathcal{B}^{-\infty}.
\end{equation}

\subsection{Second  quantization of classical super observables} 
Following  the chain rule , we define the odd and directional complex derivatives
\begin{equation}
\label{ }
\partial_{z_1}\ =\ \partial_{\varpi z_1}\varpi^\dag, \quad
\partial_{z_1^*}\ =\ \partial_{z_1^*}\varpi^\dag.
\end{equation}
These \emph{analytic} odd directional derivatives coincide with the left and right \emph{algebraic} fermionic derivatives from \textsc{berezin}\cite{Berezin-65}. 

Together  with the even directional derivatives $\partial_{z_0}$ and 
$\partial_{z_0^*}$ they define the  directional super derivatives $\partial_{z}\equiv\partial_{z_0}+\partial_{z_1}$ and $\partial_{z^*}\ \equiv\ \partial_{z_0^*}+\partial_{z_1^*}$.
The super annihilation operators 
\begin{equation}
\label{ }
\hat{z}^\dag: \mathcal{B}^{-\infty}\rightarrow
\mathcal{B}^{-\infty}\  \mbox{if}\ z^*\in\mathcal{H}^{\infty},\ \quad
\widehat{z^*}^\dag: \mathcal{B}^{\infty}\rightarrow
\mathcal{B}^{\infty}\ \mbox{if}\ z\in\mathcal{H}^{-\infty}.
\end{equation}
The corresponding super creation  operators are the super adjoint multiplication operators $\hat{z}$ and $\widehat{z^*}$.
The canonical super commutation relations: if $z^*\in\mathcal{H}^{\infty},\ w\in\mathcal{H}^{-\infty}$  then
\begin{equation}
[\widehat{z^*},\ \hat{w}]\ = \ 0\ =\  [\widehat{z^*}^\dag,\ \hat{w}^\dag],\ \quad\
[\widehat{z^*}^\dag,\ \hat{y}]\ =\ z^*w.
\end{equation}

\medskip
As in the bosonic case, the classical super observables are analytic functionals $\Theta\in(\mathcal{B}\otimes\mathcal{B}^*)^{-\infty}$.

The corresponding classical bosonic observable is the composition 
$\Theta^\varpi\equiv\Theta\circ\varpi^\dag$. The normal bosonic quantum operator $\Theta^\varpi(\hat{z},\widehat{z^*}^\dagger)$ has a unique super counterpart $\Theta(\hat{z},\widehat{z^*}^\dagger)$. Thus any continuous linear operator in a Gelfand super triple  has a unique normal bosonic symbol $\widetilde{\Theta}^\varpi(z^*,z)$ and the associated anti-normal symbol $\Theta^\varpi(z,z^*)$.

In particular, $\partial^\varpi_z, \partial^\varpi_{z^*}$ are the bosonic counterparts of the super directional derivatives, and $\hat{z}^\varpi, \widehat{z^*}^\varpi$ of the super multiplication operators. Then
$e^{\widehat{z^*}^\dagger}\varpi$ corresponds to
 $e^{\widehat{z^*}^\dagger}$, and $e^{\hat{z}\varpi}$ to $e^{\hat{z}}$.
 Thus 
 \begin{equation}
\label{}
\Theta(\hat{z},\widehat{z^*}^\dagger)^ \varpi\ =\ 
\varpi^\dag\Theta(\hat{z},\widehat{z^*}^\dagger)\varpi,
\end{equation}
so that the matrix elements  of the counterpart operators coincide:
 if $\Psi\in\mathcal{B}$, then
\begin{equation}
\label{eq:matrix}
\Psi^*\Theta(\hat{z},\widehat{z^*}^\dagger)\Psi\ =\ 
\Psi^{\varpi*}\Theta^\varpi(\hat{z},\widehat{z^*}^\dagger)\Psi^\varpi.
\end{equation} 

\section{Anti-normal super Feynman integral}
Here we derive the \emph{anti-normal} version of Feynman integral for 
the transformation  matrix elements of Schr\"{o}dinger super operators. In view of
(\ref{eq:matrix}), we consider the bosonic case only.

Let $\{p_{n}\}$ be a flag of \emph{finite dimensional} orthogonal projectors 
in $\mathcal{H}^\infty$ (i. e., an increasing sequence  of   projectors  which are orthogonal in
$\mathcal{H}$ and strongly converging to the  unit operator in  
$\mathcal{H}^\infty$. 
They   naturally define  the flag of finite dimensional orthogonal projectors in the Gelfand triple  $\mathbb{H}$ and, therefore,
 the corresponding flag of infinite dimensional  orthogonal projectors 
$\hat{p}_{n}$ in the  Gelfand triple $\mathbb{B}$.

Let $\widehat{H}\equiv\Theta(\widehat{z},\widehat{z^*}^\dagger)$.
Assume that 
\begin{equation}
\label{eq:geq}
\Theta(z,z^*)\geq 0
\end{equation}

The  contractions $\widehat{H}_n\equiv\hat{p}_{n}\Theta(\widehat{z},\widehat{z^*}^\dagger)\hat{p}_{n}$  are operators in 
 the Bargmann--Hida  triples $\mathcal{B}_n\equiv \hat{p}_{n}\mathcal{B}$ over the  finite-dimensional Hermitian spaces  
 $\mathcal{H}_n\equiv p_{n}\mathcal{H}$. 

Moreover $\Theta(p_nz,p_nz^*)$ is  the  antinormal symbol of $\widehat{H}_n$. 
(By (\ref{eq:normal}), this is straightforward for normal symbols, and then, by (\ref{eq:heat}) for antinormal as well.)

Note,  we have identified  the finite-dimensional spaces $\mathcal{H}_n$  with the Gelfand triples $\mathcal{H}_n\subset \mathcal{H}_n\subset \mathcal{H}_n$. There Minlos Gauss measure is the standard   Gauss measure on $\mathcal{H}_n$, so that $\mathcal{B}_n$ are (unbounded) operators  on the Hilbert spaces 
$\mathcal{B}_n^0$ (see textsc{berezin}\cite{Berezin-71})
with the dense domains $\mathcal{B}_n^\infty$. 

By (\ref{eq:geq}),   $\widehat{H}_n$ are positive definite symmetric operators on the Hilbert spaces $\mathcal{B}_n^0$ with the dense domains $\mathcal{B}_n^\infty$. They have Friedrichs selfadjoint extensions which are denoted again as $\widehat{H}_n$. 

Now the  transition amplitudes  in $\mathcal{H}_n^\infty$ are
\begin{equation}
\label{ }
\langle p_n z_t|\hat{p}_nz_0\rangle\ =\  e^{p_{n}z_0\,*}
e^{-it\widehat{H}_n}e^{p_nz_0},
\end{equation}
As in \textsc{klauder-scagerstam} \cite[pp.69-70]{KS},  consider  the strongly differentiable   family of operators 
$\widehat{A}_{n,\tau},\ 0\leq\tau\leq t,$ in  $\mathcal{B}$ 
\begin{equation}
[\widehat{A}_{n,\tau}\Psi](z_0^{*})= \int\!dz^{*}dz\, e^{-z^{*}z}
e^{z_0^{*}z}e^{-i\Theta_n(z^{*},z)\tau}\Psi(z^{*})
\end{equation}
Since $|e^{-i\Theta_n(z^{*},z)t}|=1$, the  operator norms   $\|\widehat{A}_{n,\tau}\|\leq 1$ in $\widehat{H}_n$. 

Besides, the strong $t$-derivative $(d/dt)\widehat{A}_{n,\tau}(0)=\widehat{H}_n$ on the exponential states.  Then, by the Chernoff's product theorem (see \textsc{chernoff}\cite{Chernoff}), the evolution operator 
\begin{equation}
e^{-i\hat{H}_n}=\lim_{N\rightarrow \infty}[\widehat{A}_{n,t/N}]^{N}.
\end{equation}
Its  kernel   is the kernel  contraction of the kernels of the factors
 \begin{equation}
 \label{eq:approx}
\int\prod_{j=1}^{N}\!dz_{j}^{*}dz_{j}\,
\exp\sum_{j=0}^{N}\Big[(z_{j+1}-z_{j})^{*}z_{j} -
it\Theta_n(z_{j}^{*},z_{j})/N\Big],
\end{equation}
where $z_{N+1}= z_t,\ z_{0} = z_0)$.

Thus the  amplitude $e^{p_{n}z_t^*}e^{-it\widehat{H}_n}e^{p_nz_0} $ is 
the $N$-iterated Gaussian integral over $\mathcal{H}$ which, by the Fubini's theorem, is equal to  the $N$-multiple  Gaussian integral over 
$\mathcal{H}^{N}$. 

In the notation $\tau_{j}=jt/N,\ z_{\tau_{j}} = z_{j},\ 
j=0,1,2,\ldots,N$,
and  $\Delta \tau_{j}= \tau_{j+1}-\tau_{j}$, the  multiple integral  is
\begin{equation}
\int\prod_{j=1}^{N}\!dz_{\tau_{j}}^{*}dz_{\tau_{j}}\:
\exp i\sum_{j=0}^{N}\Delta t_{j}\left[-i(\Delta z_{\tau_{j}}/
\Delta \tau_{j})^{*} z_{\tau_{j}}\rangle - 
\Theta_n(z_{\tau_{j}}^{*},z_{\tau_{j}})\right].
\end{equation}
 Its limit at $N=\infty$  is a rigorous mathematical  definition 
 of the heuristic   antinormal Feynman   integral
\begin{equation}
\label{eq:Feynman}
\int_{z_0}^{z_t}\prod_{0< \tau < t}\!dz_{\tau}^{*}dz_{\tau}\:
 \exp \int_{0}^{t}d\tau
\left[(\partial_{\tau}z_{\tau}^{*}) z_{\tau} - i \Theta_n(z_{\tau}^{*},z_{\tau})\right]
\end{equation}
over  classical histories $z_{\tau}$ between  $z_0$ and $z_t$
in $\mathcal{H}_n$.

Since the quantum  amplitude 
\begin{equation}
\label{ }
\langle z_t^*|z_0\rangle\ =\ \lim_{n\rightarrow\infty}\
\langle p_nz_t^*|p_nz_0\rangle,
\end{equation}
it is equal to the  iterated limit of (\ref{eq:approx}) as $N\rightarrow\infty$
is followed by $n\rightarrow\infty$. That iterated limit is  a rigorous mathematical  definition  of the heuristic   antinormal Feynman   integral for the amplitude
$\langle z_{t}|z_0\rangle$
\begin{equation}
\label{}
\int_{z_0}^{z_t}\prod_{0< \tau < t}\!dz_{\tau}^{*}dz_{\tau}\:
 \exp \int_{0}^{t}d\tau
\left[(\partial_{\tau}z_{\tau}^{*}) z_{\tau} - i \Theta(z_{\tau}^{*},z_{\tau})\right]
\end{equation}
over  all classical histories $z_{\tau}$ between  $z_0$ and $z_t$
in $\mathcal{H}^\infty$.

\emph{The non-negativity condition of the  anti-normal symbol may be replaced just by  its boundedness from below}.

\bibliographystyle{amsalpha}

\end{document}